\documentstyle[12pt,prep]{article}
\begin{document}
\pagestyle{empty}
\begin{flushright}
{CERN-TH.7309/94}\\
\end{flushright}

\vspace*{5mm}
\begin{center}
\section*{
 \boldmath{$SU(3)$} breaking
in \boldmath{$\bar{B}/D\rar Kl\bar \nu$} decays \\
and value of
\boldmath{$V_{cd}/V_{cs}$}}
\vspace*{1.5cm}
{\bf S. Narison} \\
\vspace{0.3cm}
Theoretical Physics Division, CERN\\
CH - 1211 Geneva 23\\
and\\
Laboratoire de Physique Math\'ematique\\
Universit\'e de Montpellier II\\
Place Eug\`ene Bataillon\\
F-34095 - Montpellier Cedex 05\\
\vspace*{2.0cm}
{\bf Abstract} \\ \end{center}
\vspace*{2mm}
\noindent
We determine the $SU(3)$-breaking effects in the
$\bar{B}/D \rar Kl\bar \nu$
decays from the
ratio of the $\bar{B}/D\rar Kl\bar \nu$ over the $\bar{B}/D \rar \pi
l\bar \nu$
semileptonic form factors using a {\it semi-analytic} expression
based on QCD spectral sum rules. We obtain
$f_+^{\bar{B}\rar K}(0)/f_+^{\bar{B} \rar \pi}(0)$ = $1.007\pm 0.020$ and
$f_+^{D\rar K}(0)/f_+^{D \rar \pi}(0)$ = $1.102 \pm 0.007$, which
combined with the present CLEO data,
leads to $V_{cd}/V_{cs}$ = $ 0.323 \pm 0.058$ and then, to:
$\vert V_{cs}\vert = 0.72\pm 0.10$.
The prediction on the ratio of mixing angle is (surprisingly)
much higher than
the value $0.226 \pm 0.004$
derived from the unitarity constraint on the CKM matrix, while the
corresponding value of $ V_{cs}$ is lower. These results
can open a window for new
physics, though
a stringent conclusion needs data with much higher statistics (from e.g
tau-charm factory)
or/and alternative QCD-based estimates of the ratio of
the semileptonic form factors.
\vspace*{2.0cm}
\noindent


\begin{flushleft}
CERN-TH.7309/94 \\
\today
\end{flushleft}
\vfill\eject
\pagestyle{empty}

\setcounter{page}{1}
\pagestyle{plain}

\section{Introduction}
Encouraged by the success of QCD spectral sum rules (QSSR)
for predicting
the form factors of the semileptonic heavy-quark processes using
vertex sum rules, we shall examine in this paper  the size of the
$SU(3)$ breaking in the simple but important case of the $\bar{B}/D$
semileptonic decays into $K/\pi$. Analogously to the case of the
$B \rar K^* \gamma$ \cite{SN1},
 we shall work with the ratio of form factors,
which is only sensitive to the $SU(3)$-breaking parameters and which
is expected to be
 more precise than a direct estimate of the absolute value usually
done in the literature,
 due to the cancellation of systematic uncertainties in the ratio.
For this purpose, we shall consider the
form factors $f_+^{H\rar P}$, defined in the standard way as:
\beq
\la P(p')\ve
\bar u\gamma_\mu b\ve H(p)\ra =
f_+^{H\rar P}(p+p')_{\mu}+f_-^{H\rar P}(p-p')_\mu ,
\eeq
where $H \equiv B,D$ and $P \equiv \pi,K$.
Then, we shall consider the three-point function:
\beq
V(p,p',q) = -\int d^4x \int d^4y\, \mbox{exp}
(ip'x-ipy) \la0\ve TJ_L(x)O(0)J_b(y)\ve
0\ra,
\eeq
whose Lorentz decompositions are analogous to the previous hadronic
amplitudes. Here
$J_L\equiv (m_u+m_q) \bar u i\gamma_5 q~(q\equiv d,s)$
 is the bilinear quark current having the quantum
numbers of the $\pi/K $ meson; $J_Q \equiv (M_Q +m_d)
\bar d i\gamma_5 Q~(Q\equiv b,c)$ is
the quark current associated to the $\bar{B}/D$-meson;
$O\equiv \bar b\gamma_\mu
q$ is the charged weak
current for the semileptonic transition.
The vertex function obeys the double dispersion relation:
\beq
V(p^2,p'^2,q^2)= \frac{1}{4\pi^2}\int_{M^2_b}^{\infty}\frac{ds}{s-p^2}
\int_{0}^{\infty}\frac{ ds'}{s'-p'^2} \,
\mbox{Im}\, V(s,s',q^2)+...
\eeq
\section{The \boldmath{$\bar B \rar \pi /K$} semileptonic decay }
As already emphasized in \cite{SN1}, we shall
work with the HSR introduced in \cite{SN2}:
\bea
\cal {H}(n,\tau) &\equiv &
\frac{1}{n!}\ga\frac{\partial}{\partial p^2}\dr^n
_{p^2=0}\cal{L}\ga V(p^2,p'^2,q^2) \dr  \nnb \\
&=&
\frac{1}{\pi^2}\int_{M^2_b}^{\infty}\frac{ds}{s^{n+1}}
\int_{0}^{\infty} ds' \, \mbox{exp}(-\tau' s')
\mbox{Im}\, V(s,s',q^2).
\eea
($\cal{L}$ is the Laplace transform operator) for the $B$-meson as
the sum rule has a good behaviour in the large-$M_b$ limit.
In order to come to observables, we
insert intermediate states between the charged weak
and hadronic currents in (2),
while we smear the higher-states effects with
 the discontinuity of the QCD
graphs from a threshold $t_c$ ($t'_c$) for the heavy (light)
mesons. Therefore, we have the FESR:
\bea
\cal{H}_{res}& \simeq& 2C_P f_B \frac{f_+^{B\rar P}(q^2)}{M_B^{2n}}
\mbox{exp} \, (-M^2_L\tau)
\nnb \\
&\simeq&
 \frac{1}{4\pi^2}\int_{M^2_b}^{t_c}\frac{ds}{s^{n+1}}
\int_{0}^{t'_c} ds' \, \mbox{exp}(-\tau s')
\mbox{Im}\, V_{{PT}}(s,s',q^2) + {NPT}.
\eea
$PT~ (NPT)$ refers to perturbative (non-perturbative)
contributions; $C_P \equiv f_P M_P^2$ for light pseudoscalar mesons;
$M_P$ is the light-meson mass; the
decay constant is normalized as:
\beq
 (m_q+M_Q)\la 0\ve \bar q (i\gamma_5)Q\ve P\ra= \sqrt 2 M^2_Pf_P;
\eeq
$f_+^{B\rar P}(q^2)$
is the form factor of interest.
Although we
shall not exploit the fact that $M_b$ is large, we shall work,
for convenience, with the
non-relativistic energy parameters $E$ and $\delta M_{(b)}$:
\beq
s \equiv (M_b+E)^2 ~~~~~~~~~\mbox{and}~~~~~~~~~
\delta M_{(b)} \equiv M_B-M_b,
\eeq
where, as we saw in the analysis of the two--point correlator,
the continuum energy $E_c$ is\cite{SN3}:
\bea
E^D_c &\simeq& (1.08 \pm 0.26)~\mbox{GeV} \nnb \\
E^B_c &\simeq& (1.30 \pm 0.10)~\mbox{GeV} \nnb \\
E^{\infty}_c &\simeq& (1.5 \sim 1.7)~\mbox{GeV}.
\eea
We shall use the following values of the quark masses \cite{SN4}:
\bea
\bar{m}_s(1~\mbox{GeV}) &=& 159.6\pm 8.5~\mbox{MeV} \nnb \\
M_c(p^2=M^2_c)&=& 1.46 \pm 0.05 ~\mbox{GeV}\nnb \\
M_b(p^2=M^2_b)&=& 4.59 \pm 0.04 ~\mbox{GeV}
\eea
and of the condensates \cite{SN4}:
\bea
\la \bar qq \ra (q^2)& = &-(188.9\pm 6.6 ~\mbox{MeV})^3
\log{(\sqrt{-q^2}/\Lambda)}^{4/9}
\nnb \\
{\la \bar ss \ra}/{\la \bar qq \ra}& = & 0.6 \pm 0.1 \nnb \\
\Lambda_5 &=& 180 \pm 80~\mbox{MeV}
\eea
We shall use the QCD expression
 of the vertex function obtained in \cite{OVI}
to leading order in $\alpha_s$ and including
the dimension-5 condensate contributions.
The previous form factor defined in (1) has been
{\it numerically} estimated in a number of papers,
while its {\it semi-analytic} form has been proposed in \cite{SN1}. For
the kaon, an analogous expression can be derived with the inclusion
of the $SU(3)$-breaking terms. It reads:
\bea
f^{\bar{B} \rar K}_+(0)
\simeq &-&
\frac{(m_u+m_s)\la \bar qq \ra}{4f_K m^2_K}\mbox{exp}(m^2_K\tau')
\frac{1}{f_B} \ga\frac{M_B}{M_b}\dr^{2n}  \nnb \\
&\times &\aga \ga 1+\frac{\bar m_s}{M_b}
+\delta^{(5)}_s
\dr ~\mbox{e}^{-\bar{m}^2_s\tau'}
+\frac{\cal{I}^B_K}{M^2_b} \adr ,
\eea
for a pseudoscalar quark current describing the kaon;
$\cal{I}^B_K $ is the spectral integral coming from the
perturbative graph and can be parametrized as:
\beq
\cal{I}^B_K \simeq  \cal{I}^B_\pi
\aga 1-\frac{\bar{m}_s }{M_b}+(1.7\pm 0.1)
\ga\frac{\bar{m_s}}{E_c}\dr +... \adr,
\eeq
where $ \cal{I}^B_\pi \simeq 5.3 \pm 0.6$ is the perturbative
integral in the chiral limit. It
indicates
that at $M_b=4.6$ GeV, the perturbative
contribution, although large, still
 remains a correction compared with the light-quark
condensate term; $\delta^{(5)}_s$ is the correction due to the
dimension-5 condensate and reads:
\beq
\delta^{(5)}_s \simeq
\delta^{(5)} \aga 1+\frac{\bar{m}_s}{M_b} (-0.8\sim 0.3)+... \adr
\eeq
with:
\beq
\delta^{(5)} \simeq -\tau'M^2_0 \aga \frac{(n+1)}{3}+
\frac{\tau'^{-1}}{4M^2_b}(n^2+3n+4) \adr .
\eeq
We eliminate the systematic errors due to the value of the quark masses
by introducing the  2-point pseudoscalar sum rules used (successfully)
in the estimate of the sum of the light-quark running
masses \cite{SNB}. Then,
we deduce the ratio:
\beq
\frac{(m_u+m_s)}{(m_u+m_d)}
\simeq \frac{f_K m^2_K}{f_\pi m^2_\pi}~{\mbox e}^{-\frac{\tau_2}{2}
\ga m^2_K -m^2_\pi \dr}
\aga 1+\bar{m}^2_s\tau_2 +\frac{4\pi^2}{3}
\bar{m}_s\tau_2^2\la \bar qq \ra\ga 1-\frac{1}{2}~
\frac{\la \bar ss \ra}{\la \bar qq \ra}\dr \adr ,
\eeq
where the effects of the $\pi'$ and $K'$ states as parametrized
in \cite{SNB} cancel in the ratio.
A
further optimization requirement is the standard choice $\tau'=\tau_2/2$,
between the 3- and 2-point functions sum-rule variables. This procedure
eliminates (in some sense) the exponential factor artefact of the sum
rule. At the end of the day, we deduce the $SU(3)$-breaking sum rule:
\bea
R_{B} \equiv \frac{f_+^{\bar{B}\rar K}}{f_+^{\bar{B}\rar \pi}}
&\simeq &
\aga 1+\bar{m}^2_s\tau_2 +\frac{4\pi^2}{3}
\bar{m}_s\tau_2^2\la \bar qq \ra\ga 1-\frac{1}{2}~
\frac{\la \bar ss \ra}{\la \bar qq \ra}\dr \adr \nnb \\
&\times&{\aga \ga 1+\frac{\bar m_s}{M_b}
+\delta^{(5)}_s
\dr ~\mbox{e}^{-\bar{m}^2_s\tau'}
+\frac{\cal{I}^B_K}{M^2_b} \adr}\Bigg{/}
{\aga \ga 1+
\delta^{(5)} \dr
+\frac{\cal{I}^B_\pi}{M^2_b} \adr}.
\eea
The previous {\it semi-analytical} expressions indicate that in the
large-$M_b$ limit, the $SU(3)$ breakings from the 3-point function
involving the $b$-quark vanish, while the ones brought
by the light-meson mass and coupling remain as shown in (15).
A standard stability analysis in Fig. 1
shows that the stability obtained in $\tau'$ is
perfect around 0.7--0.8 GeV$^{-2}$, while the position
of the minima increases very smoothly with $n$. In the large range of
$n$=1--4, where the absolute values of the form factor stabilize, one
can deduce with a very good
confidence:
\beq
R_{B} \equiv \frac{f_+^{\bar{B}\rar K}(0)}{f_+^{\bar{B}\rar \pi}(0)}
 \simeq 1.007 \pm 0.020,
\eeq
where we have multiplied
the obtained error by a factor of $\sqrt{2}$ in order to take into
 account the effects of the unknown
higher-order terms. One should notice that the ratio $R_B$ is not
very sensitive to the changes of the input parameters in Eqs. 8) to 10).
One should also notice that the $SU(3)$-breaking effects tend to increase
$R_B$, which is very similar to the ones encountered in the cases
 of $f_{B_S}$ and $B\rar K^*\gamma$.
\section{The \boldmath{$ D \rar K/\pi $} semileptonic decay }
For
the $D$-meson, we shall work with the {\it popular} double Laplace
sum rule (DLSR), as the $c$-quark is not heavy enough to justify
the use of the HSR based on the $1/M_c$ expansion. The DLSR reads, in its
FESR form:
\beq
{\cal L}_2= \frac{1}{4\pi^2}\int_{M^2_c}^{t_c}{ds}~\mbox{exp}{-s\tau}
\int_{0}^{t'_c} ds' \, \mbox{exp}(-\tau' s')
\mbox{Im}\, V_{{PT}}(s,s',q^2) + {NPT}.
\eeq
As in the previous case, we shall work with the non-relativistic
variables introduced in (7) and the Laplace sum-rule variable:
\beq
\tau_N \equiv 2M_c\tau .
\eeq
 Similarly to (11), one can write:
\bea
f^{D \rar K}_+(0)
\simeq &-&
\frac{(m_u+m_s)\la \bar qq \ra}{4f_K m^2_K}\mbox{exp}(m^2_K\tau')
\frac{1}{f_D}\ga \frac{M_c}{M_D}\dr^2 \mbox{exp}\aga\tau_N \delta
M^{(c)}\ga 1+ \frac{\delta M_{(c)}}{2M_c}\dr \adr \nnb \\
&\times&\aga \ga 1+\frac{\bar m_s}{M_c} \dr ~\mbox{e}^{-\bar{m}^2_s\tau'}
+\delta^{(5)}_s+\frac{\cal{I}^D_K}{M^2_c} \adr .
\eea
Here, $\cal{I}^D_K $ is the spectral integral coming from the
perturbative graph and reads:
\beq
\cal{I}^D_K \simeq  \cal{I}^D_\pi
\aga 1-\frac{\bar{m_s}}{M_c}+(1.2\pm 0.4)
\ga\frac{\bar{m_s}}{E_c}\dr +... \adr,
\eeq
where $ \cal{I}^D_\pi \simeq 12.8 \pm 1.0$, which indicates
that at the charm mass $M_c=1.46$ GeV the perturbative
contribution $numerically$ dominates (although non-leading in 1/$M_c$)
over the
the light-quark
condensate term; $\delta^{(5)}_s$ is the correction due to the
dimension-5 condensate and reads:
\beq
\delta^{(5)}_s \simeq
\delta^{(5)} \aga 1+\frac{\bar{m_s}}{M_c} (1.77\pm 0.13)+... \adr
\eeq
with:
\beq
\delta^{(5)} \simeq \frac{M^2_0}{6}\aga 2\tau+
2\frac{\tau_N}{M_c}-\frac{3}{8}\tau_N^2-M_c\tau_N\tau \adr .
\eeq
As one can see from the expression of $\delta^{(5)}$, taking the
limit $M_c$ to infinity is dangerous as the QCD series blow up. This
explicit example (among others, see e.g. \cite{SN1}) shows that
using the DLSR in the large-mass limit for the heavy-to-light
processes is not appropriate.
{}From the usual stability analysis of the sum rule analogous to (16),
we deduce from Fig.2:
\beq
R_{D} \equiv \frac{f_+^{D\rar K}(0)}{f_+^{D\rar \pi}(0)}
\simeq 1.102 \pm 0.007,
\eeq
where, as before, we have multiplied the resulting error by a factor
of
$\sqrt{2}$ in order to take into account the unknown higher-order terms.

\section{Unitarity test of the CKM-matrix}
\subsection*
{\boldmath{$V_{cd}/V_{cs}$}}
Using the CLEO II \cite{CLEO} data on the semileptonic
branching ratio:
\beq
\frac{Br\ga D^+\rar \pi^0l\nu\dr}{Br\ga D^+\rar \bar{K}^0\l\nu\dr}
=(8.5 \pm 2.7 \pm 1.4)\%,
\eeq
from which one obtains:
\beq
{V_{cd}}/{V_{cs}} = (0.292\pm 0.051)R_D,
\eeq
by using the experimental fact (polar form)
that the $q^2$-dependence of the form factors cancels
in the ratio. Using our previous prediction on $R_D$,
one can deduce \footnote[1]{MARK III \cite{MARK3}
data would imply a value of
$0.25 \pm 0.15$, which is less accurate.}:
\beq
 {V_{cd}}/{V_{cs}}
= 0.322 \pm 0.056,
\eeq
which is much larger than the prediction derived
from the unitarity constraint
on the CKM matrix:
\beq
{V_{cd}}/{V_{cs}}
= 0.226 \pm 0.005.
\eeq
If the CLEO data are confirmed, then
our result is the $first$ which signals a deviation from
the CKM unitarity constraints. Therefore, it
can suggest that $V_{cs}$
 is lower (or $V_{cd}$ larger)
 than the value
expected from the unitarity of the CKM matrix; so, it can open
a window for new physics beyond the standard model.
A much more stringent test needs further accuracy on
the data and/or alternative QCD-based estimates of the ratio of
the form factors.
\subsection*{\boldmath{$V_{cs}$} and \boldmath{$ f_+^{D\rar K}(0)$}}
Let us now estimate $V_{cs}$.
One can combine the previous result on the ratio of mixing angles with
the measured value $\vert V_{cd}\vert =0.204 \pm 0.017$ as quoted in PDG
 \cite{NU}. Then, one can deduce:
\beq
\vert V_{cs} \vert =0.63 \pm 0.12,
\eeq
which is lower than the value
$\vert V_{cs} \vert =
0.9735-0.9751$ from the unitarity constraints of the CKM matrix.

\nin
One can also extract $V_{cs}$ directly
from the measured $D\rar Kl\bar \nu$ decay rate by a theoretical estimate
of the form factor $f_+^{D\rar K}(0)$. A QSSR estimate of this quantity
already exists in the literature with the value \cite{DOSCH}
 $0.60^{+0.15}_{-0.10}$,
which is slightly lower than the experimental value $0.77 \pm 0.04$
obtained by
 CLEO II \cite{CLEO2} using the value of $V_{cs}$ from unitarity.
We shall update the previous sum-rule prediction, within the strategy
discussed previously. We shall start with the expression given in (20)
by including in this sum rule and in the 2-point function in (15),
the effect of the $K'(1.5)$, which can be parametrized through its decay
constant as:
\beq
 \frac{f_{K'}}{f_K} \simeq r_K \ga\frac{M_K}{M_{K'}}\dr^2,
\eeq
where $r_K \simeq 2.6\pm 0.2$ is a theoretical estimate \cite{GASSER}.
We perform a standard stability analysis, which we show in Fig. 3. The
stability in the light-quark variable $\tau'$
manifests itself as a clear maximum,
while the possible inflexion point in the heavy quark $\tau_N$-variable
is delicate to localize. Taking into account this localization
uncertainty by considering the physically-motivated
range $\tau_N \simeq 0.3- 0.6$ GeV$^{-1}$
and including the errors induced by the input parameters
in (8)--(10), which are
mainly due to $M_c$, to the correlated value of $f_D$ and
the continuum energy $E_c$, we deduce:
\beq
f_+^{D\rar K}(0) \simeq 0.80 \pm 0.16.
\eeq
Within the accuracy of our result and the present data from CLEO II
\cite{CLEO2}, one obtains:
\beq
\vert V_{cs} \vert =0.93 \pm 0.19.
\eeq
while a conservative lower bound of 0.72 can be derived by assuming
$f_+^{D\rar K}(0) \leq 1.$ Due to the large error induced by the absolute
value of the form factor, the unitarity test is less stringent than
the one from the ratio of form factors. By combining the two previous
results, one obtains:
\beq
\vert V_{cs} \vert =0.72 \pm 0.10.
\eeq
A sharp confirmation
of our result needs accurate data which can be obtained in a tau-charm
factory machine and/or alternative QCD-based estimates of the ratio
of form factors.
An analogous analysis
can also be pursued for the $D\rar K^*(\rho) l\bar \nu$
semileptonic decays, although the analysis is much more involved as we
have there three relevant form factors.
We plan to come back to this point in a
future project.
\vfill \eject
\section*{Acknowledgements}
Conversations with Gustavo Branco and Toni Pich have been appreciated.
\section*{Figure captions}

\nin
{\bf Fig. 1}~~~~ $\tau'$- and $n$-dependences  of the ratio $R_B$
of the $\bar{B} \rar K/\pi $ form factors.
\vspace*{0.5cm}

\nin
{\bf Fig. 2}~~~~ $\tau'$- and $\tau_N$-dependences
of the ratio $R_D$ of the $D
\rar K/\pi $ form factors.
\vspace*{0.5cm}

\nin
{\bf Fig. 3}~~~~ $\tau'$- and $\tau_N$-dependences
of the $D
\rar K $ form factor $f^{D\rar K}_+(0)$.

\end{document}